\documentclass[twocolumn,superscriptaddress,floatfix,showpacs,aps]{revtex4}
\usepackage{amsmath,amssymb,epsfig,color}
\pdfoutput=1
\usepackage{graphicx}
\usepackage{dcolumn}
\usepackage{bm}
\usepackage{color}
\usepackage{multirow}

\begin{document}

\title{Unconventional magnetism and electronic state in frustrated layered system PdCrO$_2$}
\author{Evgenia V. Komleva}
\affiliation{M.N. Mikheev Institute of Metal Physics UB RAS, 620137, S. Kovalevskaya str. 18, Ekaterinburg, Russia}

\author{Valentin Yu. Irkhin}
\affiliation{M.N. Mikheev Institute of Metal Physics UB RAS, 620137, S. Kovalevskaya str. 18, Ekaterinburg, Russia}
\affiliation{Ural Federal University, Mira str. 19, 620002 Ekaterinburg, Russia}

\author{Igor V. Solovyev}
\affiliation{National Institute for Materials Science,  MANA, 1-1 Namiki,
Tsukuba, Ibaraki 305-0044, Japan}
\affiliation{M.N. Mikheev Institute of Metal Physics UB RAS, 620137, S. Kovalevskaya str. 18, Ekaterinburg, Russia}
\affiliation{Ural Federal University, Mira str. 19, 620002 Ekaterinburg, Russia}

\author{Mikhail I. Katsnelson}
\affiliation{Radboud University, Institute for Molecules and Materials, NL-6525 AJ Nijmegen, The Netherlands}
\affiliation{Ural Federal University, Mira str. 19, 620002 Ekaterinburg, Russia}

\author{Sergey V. Streltsov}
\affiliation{M.N. Mikheev Institute of Metal Physics UB RAS, 620137, S. Kovalevskaya str. 18, Ekaterinburg, Russia}
\affiliation{Ural Federal University, Mira str. 19, 620002 Ekaterinburg, Russia}

\begin{abstract}
First-principles calculations and a model consideration of magnetically frustrated layered material PdCrO$_2$ are performed. The results on the exchange parameters are in agreement with the experimental data on  the Curie-Weiss temperature ($\theta$). We show that experimentally observed strong suppression of the N\'eel temperature ($T_N$) in comparison with the Curie-Weiss temperature  is due to three main factors. First, as expected, this is connected with the layered structure and relatively small exchange interaction along the $c$ axis. Second, deformation of the ideal in-plane 120$^{\circ}$ magnetic structure is crucial to provide finite  $T_N$ value. However, these two factors are still insufficient to explain low $T_N$ and the large frustration factor $|\theta|/T_N$. Thus, we suggest a scenario of an exotic non-Fermi-liquid state in PdCrO$_2$ above $T_N$ within the frameworks of the Anderson lattice model, which seems to explain qualitatively all its main peculiarities.


\end{abstract}

\date{\today}

\maketitle

\section{Introduction}

The metallic layered system PdCrO$_2$  possesses quite unconventional magnetic and electronic properties and demonstrates a number of puzzling (even mysterious) features. First of all, this is a rather low N\'eel temperature $T_N \approx 37$~K, while the Curie-Weiss temperature $\theta$ characterizing average exchange field is $\sim$500~K, so that frustration factor $|\theta|/T_N$ exceeds 13~\cite{Mekata1995,Takatsu}. There is also a number of anomalies in thermodynamic, spectroscopic and transport characteristics. In particular, magnetic diffuse scattering of neutrons is clearly seen  above $T_N$~\cite{Mekata1995}, which implies that the short-range spin correlations start to develop at temperatures much higher than $T_N$. Moreover, the magnetic Bragg peaks are broad even at temperatures much below $T_N$~\cite{Takatsu}. This fact implies that coherence length of the ordered moments remains finite. Below $T_N$ the conventional 120$^{\circ}$ spin structure has been observed first~\cite{Takatsu}, but later more detailed investigations found some deformations of this order~\cite{Takatsu-2014,Le-2018}.

The magnetic specific heat in PdCrO$_2$ shows a critical behavior that extends in an unusually wide temperature range above $T_N$. Interestingly, the critical exponents do not match with the exponents of the standard models, and they are also strongly asymmetric above and below $T_N$. As for the transport properties, a sub-linear temperature dependence of the electrical resistivity (with the exponent about 0.4) above $T_N$ is observed~\cite{Takatsu}. Such a behavior is quite different from what we typically have in conventional magnetic metals.  The magnetic entropy at $T_N$  is 3.9 J/mol-K~\cite{Takatsu}. This value is rather small being only one third of the expected entropy for a system with $S = 3/2$ localized spins (Cr is $3+$ with electronic configuration $3d^3$), $R \ln(2 S +1 )$ = 11.5 J/mol-K (with $R$ being the universal gaseous constant).  This again stresses presence of strong short-range spin correlations at temperatures much above $T_N$. The same conclusion was made based on an analysis of magnetotransport, namely, thermoelectric power in magnetic field~\cite{Hussey1} and anisotropic magnetoresistance~\cite{Hussey2}. 

To describe qualitatively these anomalous properties, the model of completely localized Cr spins (which corresponds to the $s-d$ exchange model~\cite{Vonsovsky,Yosida} when treating the electronic characteristics) was used in Refs.~\cite{Hussey1,Hussey2}. However, the situation can be more complicated. First, in a very recent spectroscopic study~\cite{Sunko2020} strong enough hybridization between Cr $d-$sites and metallic electrons of Pd has been observed. Also, first-principle dynamical mean field (DMFT) calculations~\cite{Lechermann} show that, despite Cr $d-$electron subsystem is strongly correlated and lies on the insulating side of Mott transition they are pretty far from the atomic limit assumed in Refs.~\cite{Hussey1,Hussey2}. Last not least, it follows from theory of quasi-two-dimensional magnets~\cite{IKK1999} that so high ratio $|\theta|/T_N$ corresponds to enormously strong anisotropy of exchange interactions which may be difficult to expect in the systems with three-dimensional metallic Fermi surface clearly seen at low temperatures~\cite{Hussey2} and therefore with (supposedly) RKKY type of exchange interactions~\cite{Vonsovsky,Yosida}. From a general point of view, it is much easier to expect a very large anisotropy due to a strong suppression of interlayer hopping and interactions in exotic (non-Fermi-liquid) phases of strongly correlated systems~\cite{Anderson,Wen1,Wen2}, especially, in combination with magnetic frustrations~\cite{Vojta}.

In the present work, we discuss the applicability of the Heisenberg and Kondo-lattice model to describe magnetism of PdCrO$_2$. In particular, we calculate from the first principles in-plane and out-of-plane exchange interactions and show that their anisotropy is far from being large enough to explain the observed ratio of $|\theta|/T_N$ within the Heisenberg model. Tiny single ion anisotropy also useless in solving this puzzle. 

The paper is organized as following. In Sec.~\ref{CalcDetails} we present  details of the computation methods. In Sec.~\ref{DFT-results} the results of density functional theory (DFT) calculations of electronic structure and exchange parameters are presented. These parameters are used to calculate various magnetic characteristics within the Heisenberg model and to demonstrate that they cannot explain the observed value of the N\'eel  temperature $T_N \approx 37$ K. In Sec.~\ref{sd} we discuss qualitatively electronic properties and the applicability of more itinerant Anderson-lattice model to the system under consideration.

\section{Computational details \label{CalcDetails}}
 
Crystal structure of PdCrO$_2$ is described by the $R\bar3m$ (166) space group. The lattice parameters were taken from Ref.~\cite{Shannon} ($a$ = 2.930 \AA~ and $c$ = 18.097\AA), while atomic positions from Ref.~\cite{Doumerc}. In order to estimate exchange interaction parameters we used the total energy calculations, which were performed for the unit cell consisted of twelve formula units.
 
We used density functional theory (DFT) within the generalized gradient approximation (GGA) \cite{GGA} taking into account strong Coulomb correlations via the GGA+U method \cite{GGAU}.  On-site Hubbard $U$ repulsion parameter was taken to be 3-4 eV and Hund's exchange $J_H$ = 0.7 eV. Similar values were successfully used for description of electronic and magnetic properties of various chromium oxides \cite{CrO2,CaCrO3} including previous DFT+DMFT calculations of PdCrO$_2$~\cite{Lechermann}.

Electronic structure GGA+U calculations were performed in the Vienna ab initio simulation package (VASP) \cite{VASP} with exchange-correlation potential chosen as proposed in Ref.~\cite{ECP}. The plane-wave energy cutoff was chosen to be 500 eV. The {\it k-}space integration was performed by the tetrahedron method and the density of the {\it k-}mesh we used was 4$\times$4$\times$4. The convergence criterion for the total energy was chosen to be 10$^{-5}$ eV.

In order to find exchange parameters $J_{ij}$ of the classical Heisenberg model, which was written in the following form:
\begin{equation}
\label{Hamilt}
H = \sum_{i>j}J_{ij}\bf{S}_i\bf{S}_j,
\end{equation}
where $i$ and $j$ numerate lattice sites, we used the total energy method as realized in the $JaSS$ code \cite{JaSS}.

We took advantage of the Luttinger-Tisza method \cite{Luttinger1946} to find the wave vector $\bf Q$ corresponding to the magnetic ground state and then used it along with the isotropic exchange parameters to estimate the N\'eel temperature. There are different options how this can be done for quasi-two-dimensional and frustrated systems with a low ordering temperature and strong short-range order above it. It is convenient to use various versions of the spin-wave theory (SWT) which include self-consistent SWT, linear SWT, and Tyablikov theory \cite{kat97,IKK1999,kat07}. The spin-wave Tyablikov approximation corresponds to the large-$S$ case of SWT (see Eqs. 2.34, 2.38, 2.43 in Ref.~\cite{kat07}). Explicit formulas for $T_N$ in the case of the spiral spin configurations  \cite{Schmidt2013,Schmidt2017} can be written for arbitrary spin value as
\begin{equation}
\label{TNeel}
T_N=\frac{1}{2}{S^2}\left(\frac{1}{N}\sum_{\bf{q}}\frac{A_{\bf{q}}}{E^2_{\bf{q}}}\right)^{-1},
\end{equation}
where $N$ is the number of $q$-points and the spin-wave dispersion $E(\bf{q})$ is expressed via coefficients $A_{\bf{q}}$ and $B_{\bf{q}}$ in the standard way
\begin{eqnarray}
\label{SWT-spectra}
E_{\bf{q}}&=&\sqrt {A_{\bf{q}}^2 -B_{\bf{q}}^2}, \\
A_{\bf{q}}&=&J_{\bf{q}}+\frac{1}{2}\left(J_{\bf{q}+\bf{Q}}+J_{\bf{q}-\bf{Q}}\right) -2J_{\bf{Q}}, \\
B_{\bf{q}}&=&J_{\bf{q}}-\frac{1}{2}\left(J_{\bf{q}+\bf{Q}}+J_{\bf{q}-\bf{Q}}\right),
\end{eqnarray}
where $J_{\bf{q}}$ are corresponding Fourier transforms of exchange parameters $J_{ij}$. 

It should be noted that these approximations are not valid if $T_N$ is not small as compared to its mean-field value, so that spin-wave theory does not work in this case. In such a situation, we can use the high-temperature Tyablikov approximation  \cite{pl} which takes into account non-Bose commutation relations of spin operators and provides an interpolation to the mean-field approximation:
\begin{equation}
\label{TNeel1}
T_N=\frac{1}{2}\frac{S(S+1)}{3}\left(\frac{1}{N}\sum_{\bf{q}}\frac{A_{\bf{q}}}{E^2_{\bf{q}}}\right)^{-1}.
\end{equation}
This result differs from Eq.~(\ref{TNeel}), if $S>1/2$. Also, an account of renormalization factors of self-consistent SWT and field-theoretical corrections can modify the results of linear spin-wave theory by a factor about 1.5 \cite{kat97,IKK1999,kat07}.
However, such uncertainties will be not too important for our conclusions.
\begin{figure}[t!]
	\centering
	\includegraphics[width=.45\textwidth]{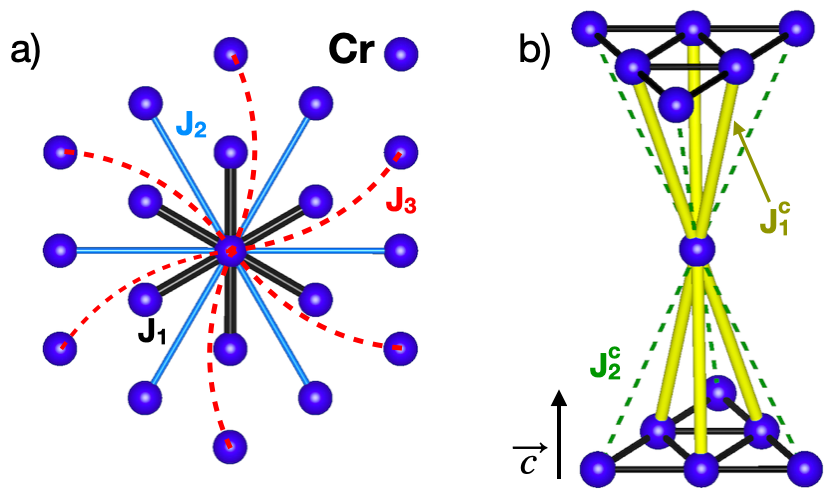}
	\caption{(a) In-plane exchanges paths $J_1$, $J_2$, $J_3$ and (b) out-of-plane $J_1^c$, $J_2^c$ considered in the article.}
\label{Exchanges_pic}
\end{figure}

The summation in \eqref{TNeel} was performed using the 500$\times$500$\times$500 mesh over the unit cell of the reciprocal space given by the following vectors: $\textbf{b}_1=(\frac{2\pi}{a}, -\frac{2\pi}{\sqrt{3}a}, -\frac{2\pi}{3c})$, $\textbf{b}_2=(0, \frac{4\pi}{\sqrt{3}a}, -\frac{2\pi}{3c})$, and $\textbf{b}_3=(0, 0, \frac{2\pi}{c})$. 

\section{DFT results: exchange parameters and  N\'eel temperature\label{DFT-results}}

Previous DFT calculation of exchange parameters reported in Ref.~\cite{Singh} unfortunately did not take into account strong Coulomb correlations, which were recently shown to be important for PdCrO$_2$~\cite{Lechermann,Sunko2020}. Also, only two in-plane isotropic exchange parameters were calculated there, namely, between the nearest and next-nearest neighbors. Interestingly, {\it ab initio} calculations for other Cr-based delafossites, MCrS$_2$ systems with M=Li, Na, K, Ag, and Au, demonstrated that exchange interactions between third nearest neighbors are not small~\cite{Alex}. Moreover, it was shown to control the magnetic structure of these materials. It motivated us to consider all these exchange paths (and moreover two out-of-plane ones, see Fig.~\ref{Exchanges_pic}) in our calculations and check whether the same situation realizes in PdCrO$_2$.

\begin{figure}[t!]
	\centering
	\includegraphics[width=.45\textwidth]{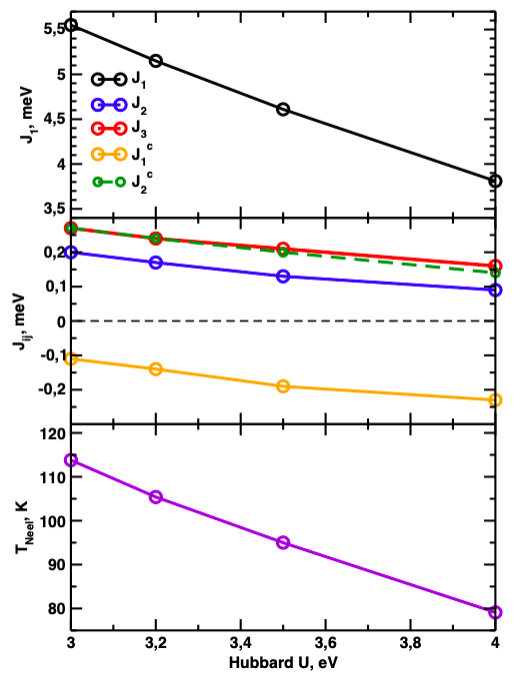}
	\caption{Dependence of the calculated isotropic exchange parameters and N\'eel temperature on Hubbard $U$. $J_1$, $J_2$, and $J_3$ are first, second, and third nearest neighbor exchange couplings (in the $ab$ plane), while $J_1^c$ and $J_2^c$ stand for exchanges between triangular planes.}
\label{JandT}
\end{figure}

Table \ref{Exchanges} summarizes our results obtained within the GGA+U calculations for various $U$, while Fig.~\ref{JandT} illustrates them. First of all, one might see that while the exchange constants depend on Hubbard $U$, the results do not change qualitatively if $U$ is varied within reasonable limits.

\begin{table}[t]
 \begin{tabular}{ccccc}  
 \hline
 \hline 
 $J_{ij}$ & $U$ = 3 eV & $U$ = 3.2 eV  &$U$ = 3.5 eV & $U$ = 4.eV \\
 \hline
$J_1$    & 5.55  & 5.15 &4.61 & 3.81\\
$J_2$    & 0.20 & 0.17 &0.13 & 0.09 \\  
$J_3$    & 0.27 & 0.24 &0.21 & 0.16 \\ 
$J_1^c$ & -0.11 & -0.14 &-0.19 & -0.23  \\
$J_2^c$ & 0.27 & 0.24 &0.20 & 0.14 \\
 \hline
 $\theta_{CW}$ &  -539 & -493 & -431 & -346 \\
 \hline
 \hline
 \end{tabular}
	\caption{Calculated in the GGA+U approximation parameters of the isotropic exchange interactions (in meV) for various values of Hubbard $U$ ($J_H$ = 0.7 eV). $J_1$-$J_3$ are in-plane exchange paths, while $J_1^c$ and $J_2^c$ correspond to the exchange interaction between the first and the second out-of-plane neighbors. In last row Curie-Weiss temperatures (in K) recalculated from these exchange parameters are presented.}
		\label{Exchanges}
\end{table}

It is revealing that similarly to the results of Ref.~\cite{Alex} in our case interaction between the third in-plane neighbors ($J_3$) also appears to be of the same order as for the second ($J_2$) ones. This, $J_3$, coupling was argued to occur by means of super-super exchange mechanism via $p$ orbitals of two adjacent ligands~\cite{Alex,UFN}. While in MCrS$_2$ it is even larger than the exchange interaction between nearest neighbors (by absolute value), in our case much less spatially extended O $2p$ orbitals (comparing to S $3p$) make this coupling less efficient. Also additional mechanism - RKKY interaction - in metallic PdCrO$_2$ can modify exchange coupling in our case. The $U$-dependence of the interaction between the third in-plane neighbors appears to be of the same order as for the second ones. 

Strong antiferromagnetic exchange coupling between the first nearest neighbors suggests the 120$^{\circ}$ spin ordering, while both antiferromagnetic $J_2$ and $J_3$ frustrate it. We used  the Luttinger-Tisza method \cite{Luttinger1946} to determine the magnetic ground state for calculated exchange parameters and found that it does correspond to the 120$^{\circ}$ structure  ($\textbf{Q}=(\frac{2\pi}{3a}, \frac{2\pi\sqrt{3}}{3a}, ...)$) if we consider purely $2D$ triangular lattice. An account of both interlayer $J^c_1$ and $J^c_2$ exchange interactions leads to a slightly different in-plane magnetic structure so that an almost 120$^{\circ}$ structure is realized. The new $\textbf{Q}=(\frac{2\pi}{3a}+\delta_x, \frac{2\pi\sqrt{3}}{3a} +\delta_y, \frac {\pi}{3c})$, where $\delta_x=0.078/a$ and $\delta_y=0.352/a$, corresponds to one $\approx$110$^{\circ}$ and two $\approx$125$^{\circ}$ in-plane angles between the magnetic moments in a one triangle. It is interesting that interlayer exchange interaction is of the order of $J_3$, with one of the out-of-plane exchange parameters being ferromagnetic. It is also important that for the ideal 120$^{\circ}$ magnetic structure the  exchange interaction between triangle planes should be zero. 

In the mean-field theory, one may recalculate the Curie-Weiss temperature 
\begin{eqnarray}
\theta_{CW} = -\frac {S(S+1)}3 J_{\bf{q}=0}.
\end{eqnarray}
Resulting values for different choices of Hubbard $U$ are summarized in Tab.~\ref{Exchanges}. We see that the best agreement with experimentally observes $\theta^{exp}_{CW} \approx -500$~K~\cite{Mekata1995} takes place for $U=3.2$ eV. We note also that  $\theta_{CW}$ in PdCrO$_2$ is much larger than what was measured in sulfides (110 K maximum \cite{Bongers1968}).

The corresponding mean-field value of the  N\'eel temperature is somewhat lower owing to frustrations and is mainly determined by intralayer exchange parameters,
\begin{equation}
T_{N}^{MF} = \frac {S(S+1)}3 J_{\bf{q}=\bf{Q}}
\end{equation}
and is about 250 K.
\begin{figure}[t]
	\centering
	\includegraphics[width=.5\textwidth]{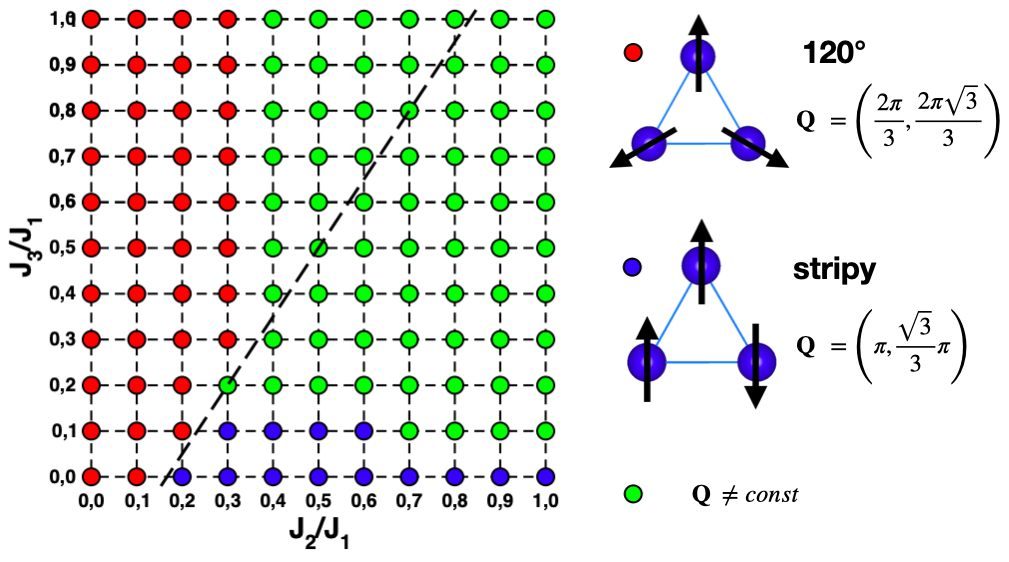}
	\caption{Calculated phase diagram ($\bf{Q}$-vector) of a $2D$ triangular lattice model with three in-plane exchange parameter. Our values of $J_2/J_1$ and $J_3/J_1$  are close to 0.03 and 0.05 respectively. In green region $\bf{Q}$-vector depends on exchange constants and changes with $J_2/J_1$ and $J_3/J_1$ ratios. Without going to details we would only like to mention that in fact there are two different phases in the green region. The dashed black line corresponds to the spin-wave instability,  vanishing $\widetilde{a}$ parameter in Eq.~\eqref{SWT-spectra}.}
\label{PD}
\end{figure}

In fact, in the purely two-dimensional situation one should have $T_N =0$ according to the Mermin-Wagner theorem. The finite interlayer exchanges  determine suppression of $T_N$ in our layered structure \cite{IKK1999}.  Since the experimental N\'eel temperature is very low, $T_N = 37$ K, we can use the linear spin-wave theory result (\ref{TNeel}) to estimate its value.
To clarify the physical picture, we consider the spin-wave spectrum near its zeros at  $\textbf{q}=0$ and $\textbf{q}=\textbf{Q}$, which lead to singularities at calculating $T_N$ according to (\ref{TNeel}). This can be done by expansions
\begin{eqnarray}
\label{SWT-spectra}
E_{\bf{q}}^2 &=& 4(4.5J_1+0.05J_2+4.5J_3)  F(\bf{q}),\\
\label{SWT-spectra}
E_{\bf{Q-q}}^2 &=& (9J_1-0.2J_2+9 J_3)  F(\bf{q}),
\end{eqnarray}
where
\begin{equation}
\begin{aligned}
F({\bf{q}})&=(1.4J_1-9J_2+6J_3) q_x^2 \\&+   (1.6J_1-9J_2+6J_3) q_y^2\\ &+  (-0.26J_1^c+0.56J_2^c)q_z^2\\ &+ (0.36J_1 - 0.1J_2 - 2.6J_3 + 0.9J_1^c +3.1J_2^c)q_xq_y\\&+
(J_1^c-1.74J_2^c)(3.1q_xq_z + 1.8q_yq_z).
\end{aligned}
\end{equation}
After diagonalization this can be written down in the new axes representation as
\begin{equation}
F({\bf{q}})= a_xq_x^2 +  a_yq_y^2 + a_z q_z^2, 
\end{equation}
where
\begin{eqnarray}
\begin{aligned}
\label{SWT-spectra}
a_x&=1.7J_1-9J_2+6J_3, \\
a_y&=1.3J_1-9J_2+6J_3, \\
a_z&=-0.26J_1^c+0.56J_2^c-1.95(J_1^c-1.74J_2^c)^2/J_1\\&\approx0.01J_1.
\end{aligned}
\end{eqnarray}
The last (approximate) value for $a_z$ is rewritten according to the previously obtained $J_i/J_1$ values.

Formally, at $a_z \rightarrow 0$ the integral (sum) in (\ref{TNeel}) is logarithmically divergent in the $z$ direction at the singularity points, and a finite value of $a_z$ provides a natural cutoff for the divergence, which defines suppression of the N\'eel temperature due to quasi-2D magnetic structure.

In the purely two-dimensional situation one returns to the 
ideal 120$^{\circ}$ structure with
\begin{equation}
\label{SWT-spectra22}
F({\bf{q}})= \widetilde{a}(q_x^2 +  q_y^2), \quad \widetilde{a} = 1.5J_1-9J_2+6J_3. 
\end{equation}
Then the $J$-dependence of the prefactor $\widetilde{a}$  demonstrates the role of frustrations of exchange interactions. For $\widetilde{a} \rightarrow 0$ (straight line in Fig.~\ref{PD}) we have softening of spin-wave spectrum, so that the zero-point spin-wave correction to the sublattice magnetization of the triangular lattice diverges and the magnetic structure becomes unstable (cf. Ref. \cite{IKK}).

The numerical integration in (\ref{TNeel}) gives N\'eel temperatures equal to 114 K ($U=3$ eV), 105 K ($U=3.2$ eV), 95 K ($U=3.5$ eV), and 79 K ($U=4$ eV), as illustrated in Fig.~\ref{JandT}.  We see that, first of all, the calculated values of $T_N$ are not too low and therefore the spin-wave theory should be applicable.  Second, as it has been explained above, the exchange parameters corresponding to $U=3.2$ eV fit the experimental Curie-Weiss temperature the best and therefore realistic estimation of N\'eel temperature, which can be obtained by the spin-wave theory is $T_N= 105$ K. This still overestimates $T_N$ in nearly 3 times. Thus our calculations demonstrate insufficiency of the localized-spin model to describe PdCrO$_2$. 

It's worth mentioning that the Tyablikov approximation (\ref{TNeel1}) for $S=3/2$ (which is appropriate for higher $T_N$ and seems to be  inapplicable in the present situation) yields the values which are lower by the factor of 5/9 and  underestimate $T_N$ in the Heisenberg model. Other versions of modified spin-wave theory \cite{kat07} can provide the values which are slightly smaller as compared to (\ref{TNeel}). However, this does not change the above conclusion.

\begin{figure}[t]
	\centering
	\includegraphics[width=.45\textwidth]{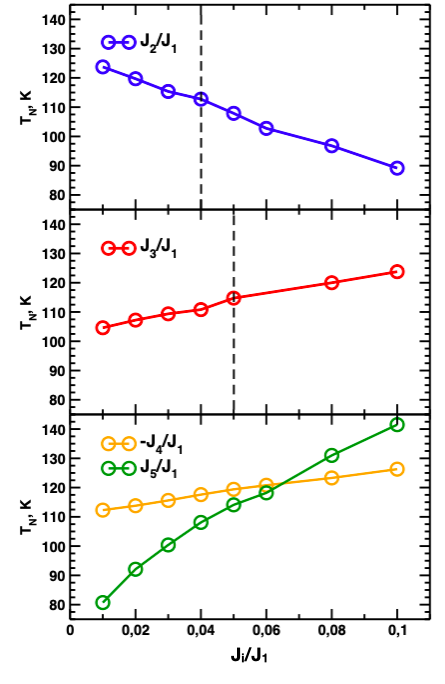}
	\caption{ Dependence of the transition temperature on the  $J_2/J_1$,  $J_3/J_1$, $|J_1^c/J_1|$, and $J_2^c/J_1$ ratios. For each presented set all the rest exchange parameters were fixed at the DFT calculated ratios obtained with the Hubbard $U$ = 3 eV (see Table \ref{Exchanges}).}
\label{TN_J}
\end{figure}

One can see from Fig.~\ref{TN_J} that $T_N$ depends appreciably on small exchange parameters $J_2$ and $J_3$, so that the role of frustrations in formation of magnetic state is considerable, but not decisive. This figure also illustrates the lowering of the transition temperature $T_N$ with the decrease of the total out-of-plane exchange parameters $J_1^c$ and $J_2^c$ thus confirming that our calculations are in agreement with the Mermin-Wagner theorem.

Finally, another factor, that is, magnetic anisotropy should be taken into account in discussions of the long-range magnetic order. The Dzyaloshinskii-Morya interaction is forbidden by symmetry and thus we computed only single-ion anisotropy, again, via total energy method, but now taking into account the spin-orbit coupling (SOC).  

The simplest ferromagnetic structure was used for this purpose. For $U = 3.0$ eV we obtained that the configuration with all spins lying in the $ab$ plane is slightly lower in energy than others, while the single-ion anisotropy constant $D$ (introduced via $D(S_z)^2$ term in the spin-Hamiltonian) is tiny $\sim 0.04$ meV and is in fact beyond the accuracy of our calculations. $D$ is significantly smaller than any of the calculated out-of plane exchange interactions, $J_1^c$ or $J_2^c$. Thus the single-ion anisotropy can not practically influence the N\'eel temperature.

\section{Discussion of electronic properties: Kondo lattice model~\label{sd}}

Anomalous transport properties of PdCrO$_2$ were discussed in Refs.~\cite{Hussey1,Hussey2} within the picture of completely localized chromium spins described by quasi-two-dimensional  Heisenberg model. However, microscopic calculations of the exchange parameters presented above clearly show that this model does not provide, at least, quantitatively correct description of the system since the ratio of in-plane to out-of-plane exchange parameters is clearly not large enough to explain the very broad range of short-range order without long-range order, that is, the experimentally observable value of $|\theta|/T_N$. We have to think therefore on alternative and probably more complicated picture.

Recent first-principle density functional plus dynamical mean-field theory (DFT+DMFT) calculation of electronic structure \cite{Lechermann} clearly showed that $3d$ electron subsystem of Cr is strongly correlated and lies in the Mott-insulator region of the phase diagram. Nevertheless, the situation is far from atomic limit, it is definitely not like in rare-earth elements \cite{Locht}. 

\begin{figure}[t]
	\centering
	\includegraphics[width=.45\textwidth]{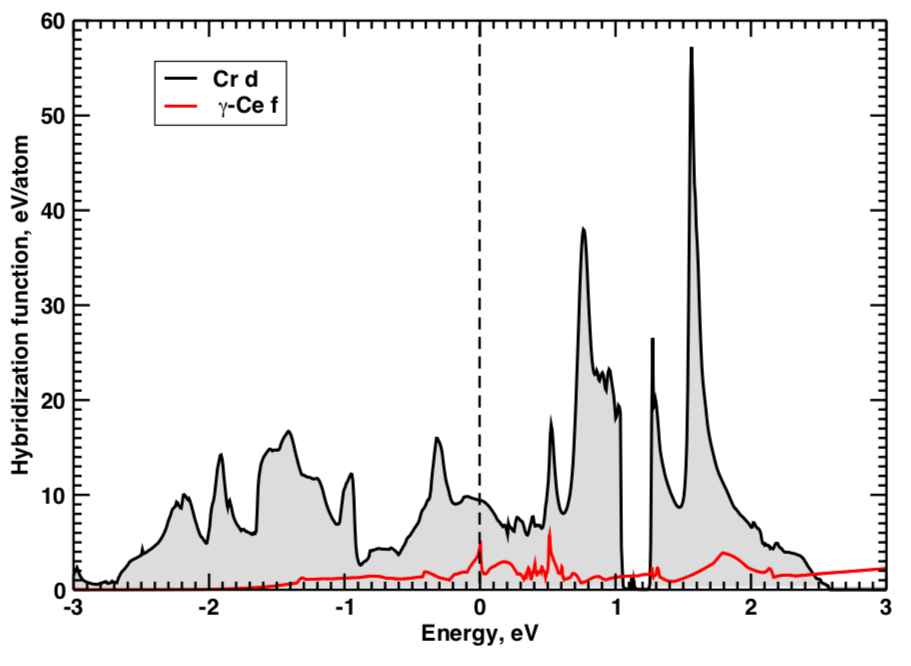}
	\caption{Hybridization function for the $d$ states of Cr atom obtained according to Ref.~\cite{Hybr_func}. For comparison the hybridization function for the $f$ states in $\gamma$-Ce is also presented~\cite{Ce}.  The Fermi level corresponds to zero energy.}
\label{Hf}
\end{figure}

Note that despite the importance of correlation effects for the electronic structure within the DFT+DMFT approach it does not mean that these effects are equally important for the value of exchange parameters. It is known (although not understood completely) that the DMFT corrections (i.e. frequency dependence of the self-energy) to the exchange parameters are typically much weaker than those for the spectral density, and one can safely use DFT or DFT+U values~\cite{Kvashnin}.

According to Ref. \cite{Sunko2020},  weak binding energy
dependence and Cr character of the reconstructed weight indicate that the spectroscopic properties of PdCrO$_2$ are essentially determined by a Kondo coupling of nearly free electrons in metallic layers, with localized electrons in a Mott insulating state in adjacent layers.

Thus, the system can be described by the $s-d$ exchange (or Anderson lattice) model. However, the situation is far from the Kondo-lattice strong-coupling heavy Fermi liquid regime (heavy-fermion situation) which occurs at rather small $s-d$ exchange coupling parameter $|I|$, so that we have an exponentially small energy scale, the Kondo temperature of order of $\exp(-1/2|I|N(E_F))$ ($N(E_F)$ is the bare density of states at the Fermi level). With the values of hybridization function as large as shown in Fig.~\ref{Hf} we are definitely far from the Kondo (or heavy-fermion) limit. Therefore the electronic specific heat is not considerably enhanced. The situation is more close to a spin liquid regime. Indeed, in the case of moderate frustration even relatively small $|I|$ results in tendency to its stabilization (the Kondo-stabilized spin liquid) \cite{Coleman,Coleman1}. At the same time, the N\'eel temperature can remain finite, although small. Below $T_N$, we have an ordered localized-moment regime, in agreement with the experimental data on PdCrO$_2$. However, with increasing $T$ we come to an exotic regime with disordered moments.

The scaling consideration of magnetic ordering formation in the Kondo lattices with usual spin-wave dynamics  \cite{IK,I20} gives as a rule a sharp  crossover with a  non-Fermi-liquid (NFL) behavior in a narrow region.
A scaling theory of the Kondo lattices with frustrated exchange interactions in spirit of the self-consistent spin-wave theory (SSWT), where spin-wave frequency does not vanish in the paramagnetic region \cite{IKK,kat07}, was presented in Ref.~\cite{I20}. This yields, depending on the model parameters,  one or two quantum phase transitions into non-magnetic  spin-liquid and Kondo Fermi-liquid ground states  with increasing the bare coupling constant. Whereas the renormalization of the magnetic moment in the ordered phase can reach orders of magnitude, spin fluctuation frequency and coupling constant are moderately renormalized in the spin-liquid phase. 

In our case we have a  different situation -- ground state moment is weakly renormalized, so that the magnetic transition is rather sharp. At the same time, strong short-range order and NFL features are observed in transport properties, so that SSWT picture is  insufficient and we have to go beyond spin-wave picture, e.g., including spinon excitations (see discussion  in Ref.~\cite{I20}).

Moreover, it seems that the simple one-parameter  scaling consideration is inappropriate,  and PdCrO$_2$ situation corresponds to  an exotic strong-coupling regime, so that a more detailed treatment of this region is required.
A description can be given in terms of the fractionalized Fermi liquid (FL$^*$) concept  \cite{Sachdev2,Sachdev1}, the non-Fermi-liquid FL$^*$ state  with deconfined neutral $S=1/2$ excitations being essentially a metallic spin-liquid state.
Here we have an  instability of the heavy Fermi-liquid  (FL) Kondo-lattice state   towards a magnetic metal where the local moments ($d$($f$)-states) are not part of the Fermi sea.

In  Ref.~\cite{Sachdev2}, the  FL$^*$  state was considered as a ground state. However, in the next paper \cite{Sachdev1} it was concluded that  with lowering temperature this state ultimately  goes to an antiferromagnetic state.
The FL$^*$  theory  includes    two distinct diverging time scales, the shorter one describing fluctuations owing to the reconstruction of the Fermi surface, and a longer one due to fluctuations of the magnetic order parameter. The intermediate time scale physics on the ``magnetic'' side is suggested to be that of a novel FL$^*$ state.  Thus, on   the magnetic side of the quantum phase transition into the FL state, there should be an intermediate temperature regime $T_N < T < T^*_{\rm coh}$ where we have  the FL$^*$ picture with the small Fermi surface which does not include $d(f)$-states (see Fig. 5 of Ref.  \cite{Sachdev1}). The presence of two  diverging length scales will influence the scaling properties of a number of physical quantities near the quantum critical point. 

This picture naturally explains incoherent character of electron motion perpendicular to Cr layers observed by angular-dependence magnetoresistance in Ref.~\cite{Hussey2}. One can simply adopt an old theory of Anderson~\cite{Anderson} suggested for high-temperature superconducting cuprates. There is a consensus now that his initial suggestion contradicts experimental data for the cuprates. Nevertheless, theoretically it is correct that the tunneling between layers is strongly suppressed and becomes incoherent, if the electronic states within the layers are not Fermi liquids and demonstrate some kind of charge-spin separation. It well may be that this idea initially suggested for the cuprates is applicable rather to PdCrO$_2$. It would be extremely interesting to check it experimentally in more detail. 

In particular, the study of optical conductivity in the direction normal to Cr layer, $\sigma_{cc}(\omega)$, can be very informative. If our picture is correct and we have more or less normal Fermi liquid below $T_N$ and incoherent state above $T_N$ one can expect some decrease of the optical spectral weight, $$S=\int_0^{\omega_0}  \sigma_{cc}(\omega) d\omega,$$ when the temperature crosses  the N\'eel  point (here $\omega_0$ is some properly chosen cutoff parameter). Importantly, $S$ is proportional to the kinetic energy of electron motion in the $c$ direction and therefore provides a direct information on the renormalization of interlayer electron hopping. Second, in the model of incoherent interlayer electron tunneling one cold expect a peculiar frequency dependence $\sigma_{cc}(\omega) \propto \omega^\alpha$ with some noninteger $\alpha$ \cite{Anderson}. 

Since the effective exchange interaction between Cr ions is indirect, most likely of the RKKY type, the suppression of the interlayer electron hopping should lead also to strong temperature dependence of effective interlayer exchange parameters. The latter can be extracted from measurements of spin wave spectra at different temperatures by inelastic neutron scattering. Note that for quasi-to-simendional magnets the spin waves are well defined in almost whole Brillouin zone up to the temperatures of the order of $|\theta|$ rather than $T_N$ \cite{IKK1999}. This kind of experiments look also very promising to solve the mystery of PdCrO$_2$. 

Another interesting consequence of our calculations is the closeness of the system to the border of stability of 120$^{\circ}$ N\'eel ground state (see Fig. 3). In Ref.~\cite{Hussey1} a giant magnetothermoelectric power was observed and explained in terms of magnon drag suppressed by magnetic field. Strong magnon drag corresponds to the regime where the rate of electron-magnon scattering is much higher than that of magnon-magnon one. For the  triangular-lattice Heisenberg model  three-magnon scattering processes are forbidden; magnetic field allows such processes and increases thereby essentially the magnon-magnon scattering rate~\cite{zhit}. Since the amplitude of the three-magnon scattering is proportional to the ratio of magnetic field to some combination of exchange energies related to the stability of the N\'eel state, relative closeness to the transition into stripy phase should enhance further the probability of these processes and  the effect of magnetic field on thermoelectric power.

\section{Conclusions}

Basing on the first-principles calculations of exchange interactions, we have demonstrated that PdCrO$_2$ cannot be described by the frustrated Heisenberg model. In particular, it is impossible to explain within this picture a very large ratio $|\theta|/T_N$ characteristic of this system. Interestingly enough, the far-neighbor exchange interactions turn out to be relevant in the determination of the magnetic ground state which shows a small deviation from the 120$^{\circ}$ magnetic structure suggested before. 

Keeping in mind also anomalous transport properties of PdCrO$_2$ we hypothesise the formation of an exotic state, possibly of FL$^*$ (metallic spin liquid) type. The issue  definitely deserves further theoretical and experimental studies. 


\section{Acknowledgements}
Authors are grateful to Yu. N. Skryabin for useful discussions.

This research was carried out within the state assignment of FASO of Russia via program ``Quantum'' (No. AAAA-A18-118020190095-4).  We also acknowledge support by Russian Ministry of Science via contract 02.A03.21.0006. The work of M.I.K. is supported by European Research Council via Synergy Grant 854843 - FASTCORR.

\end{document}